\documentclass{article}

\usepackage[final,dblblindworkshop]{neurips_2026}
\workshoptitle{AI for Science: Verification in the Age of AI Scientists}

\makeatletter
\renewcommand{\@noticestring}{%
  Accepted at the AI for Science workshop (NeurIPS 2026).%
}
\makeatother

\usepackage[utf8]{inputenc}
\usepackage[T1]{fontenc}
\usepackage{hyperref}
\usepackage{url}
\usepackage{booktabs}
\usepackage{array}
\usepackage{amsmath}
\usepackage{amssymb}
\usepackage{microtype}
\usepackage{xcolor}

\title{Independent Verification Paths Are Not Independent:\\
       A Case Study of Common-Mode Failure\\ in a Satellite Catalogue Pipeline}

\author{Fabio Rovai\\
  The Tesseract Academy\\
  London, United Kingdom\\
  \texttt{fabio@thetesseractacademy.com}}

\begin{document}

\maketitle

\begin{abstract}
A common safeguard for a data pipeline is redundant computation: derive each
published number by two routes built on different technology and refuse to exit
when they disagree. We report one such gate failing, in a cross-catalogue
integrity study of two open registers of Earth-orbiting objects. A gate
comparing a set-based Python path with SPARQL queries over the emitted RDF graph
printed \texttt{ALL CROSS-CHECKS AGREE} on seven counts. Three were wrong, one
overstated more than fourfold (932 against 220). Both paths imported the same
constants, which encoded a misreading of the source's status vocabulary, so the
error was common-mode and the gate could not see it. We give the mechanism, an
object-level ledger reconciling every figure, and three checks that go back to
the source's documentation, measured on the defective code and on its
correction. We then checked that correction against each object's phase
history, held in a source file the pipeline never read. The correction was also wrong: 42 of its 261 disagreements are
artefacts, and none of our three checks flagged them. Finally, in a controlled
replication with three pinned models and tools disabled, 72 of 75 paths
generated on request as independent checks computed the defective count, 29 of
30 even when the prompt carried the source's own definitions of the codes. The evidence is one pipeline and one
defect family. Within it, redundancy verified implementation, and the errors
that reached publication were errors of meaning.
\end{abstract}

\section{Introduction}

A pipeline that produces more numbers than its authors can trace by hand has to
check itself, and one standard way to do it is redundancy: compute every
quantity that will be published by two routes built on different technology,
compare them, and refuse to exit successfully unless they agree. The check is
cheap, mechanical, and needs no ground truth, which is why it appeals wherever
findings are generated faster than they can be audited, including in systems
that propose and test hypotheses automatically \citep{lu2024aiscientist}. The
same intuition, that agreement between routes stands in for a correctness
signal, underwrites self-consistency decoding \citep{wang2023selfconsistency}
and the use of agreement between generated code and generated tests to rank
programs \citep{chen2023codet}.

This paper is a case study of one such gate failing, in a pipeline we built. It
is a single incident, and we do not claim that it tells us how often redundancy
gates fail. What it can do is show one failure completely: the gate, the wrong
and corrected numbers, the definition that caused the error, and what did and
did not catch it.

The pipeline compares two independently maintained public catalogues of
artificial Earth-orbiting objects and emits an RDF graph of typed defect
assertions. A dual-computation gate counts each defect class twice, once
set-based in Python from the source files and once by SPARQL over the emitted
graph, and exits non-zero on any disagreement. On the run whose numbers were
published it printed \texttt{ALL CROSS-CHECKS AGREE}. Three of the seven counts
it compared were wrong, and the largest was overstated more than fourfold. Both
paths imported the same classification constants, and those constants encoded a
misreading of the source's vocabulary of flight-phase status codes. Two correct
implementations of one wrong reading agree perfectly, so the error was
common-mode and the gate was silent by construction. The multiversion
programming literature established long ago that separately developed versions
of a program do not fail independently
\citep{knight1986experimental, eckhardt1985theoretical, littlewood1989conceptual};
our case is a data-pipeline instance of that result in which the shared artifact
is not a specification but an executable set of constants.

The contributions are four, each bounded to this pipeline.
First, the mechanism, with an object-level ledger that reconciles every figure
across versions of the count (Sections~\ref{sec:failure}
and~\ref{sec:mechanism}, Appendix~\ref{app:ledger}).
Second, three checks that go back to the source's own documentation, each
implemented and run on the defective code and on its correction
(Sections~\ref{sec:caught} and~\ref{sec:gates}).
Third, a second-order failure. Checking our correction against every object's
phase history showed that it misread the same vocabulary in a different way, and that
none of our three checks had noticed (Section~\ref{sec:second}).
Fourth, an experiment on whether a code-generating model asked for an
independent verification path produces one (Section~\ref{sec:automation}).

Redundancy remains worth having; it verifies implementation. The errors that
reached publication here were errors of meaning, which two paths built on shared
definitions cannot test.

\section{Setting}

\subsection{Two registers of one population}

Artificial objects in Earth orbit are catalogued more than once. We used two
sources, both openly retrievable without credentials, both retrieved on
18~August~2026. CelesTrak's satellite catalogue \citep{celestrak2026satcat}
carried 70,292 objects. The General Catalog of Artificial Space Objects (GCAT)
\citep{mcdowell2020gcat}, maintained independently and published under CC~BY
4.0, is split across many files. The pipeline read four of them: a main
catalogue of 69,999 records covering 69,391 numbered objects, an auxiliary
catalogue of 11,962 records, a catalogue of 1,802 objects that failed to orbit,
and a catalogue of 354 numbered objects above 100,000. Two registers describing
one population is a useful setting because neither has to be treated as ground
truth: where they disagree, the disagreement itself is the finding.

\subsection{The pipeline and its gate}

The pipeline reconciles objects by catalogue number, emits an RDF graph in which
every defect is a typed node carrying its evidence, and validates that graph
against three layers of SHACL shapes \citep{shacl2017}, one shape per defect
class. On the published run it gated seven defect classes, among them entries
one register says correspond to no real object while the other still carries
them, and objects on which the two registers disagree about whether the object
is still in orbit.

Verification was deliberately over-provisioned. Three independent checks ran on
every build: a SHACL processor over all three shape layers, a second RDF engine
written in a different language that loaded and validated the same ontology, and
a dual-computation gate that recomputed every published count twice. The third
is the subject of this paper, and its structure is the obvious one:

\begin{quote}\small
\texttt{python\_side()} reads the source files and returns seven counts.
\texttt{sparql\_side()} parses the emitted graph and issues one
\texttt{SELECT (COUNT(DISTINCT ?n))} per defect class. \texttt{main()} compares
them key by key and calls \texttt{sys.exit(1)} if any pair differs.
\end{quote}

The diversity is real. One path never constructs a graph and the other never
sees the source files. One counts Python set members, the other counts distinct
SPARQL bindings over a materialised graph of 2,330,660 triples, two orders of
magnitude larger than the input. Parsing that graph takes over an hour and
roughly 5~GB of memory, so the gate is not run casually, which is part of why its
verdict carried weight. On the published run it agreed on all seven counts.

\section{The failure}
\label{sec:failure}

Table~\ref{tab:gate} reproduces the gate's output on the published run, next to
two later values for each count: the first correction, and the value after the
check against object histories described in Section~\ref{sec:second}. The gate
exited zero. Every count in this paper is a number of objects present in the
catalogues concerned, and Appendix~\ref{app:ledger} tracks each disposition
object through all three versions.

\begin{table}[t]
\centering
\small
\caption{Gate output on the published run. Every row agreed between the two
paths, and three rows were wrong. The first correction is the value published
as a fix; the full-history value is the check in Section~\ref{sec:second}.}
\label{tab:gate}
\begin{tabular}{lrrcrr}
\toprule
Defect class & Python & SPARQL & Agree & First correction & Full history \\
\midrule
PhantomEntry              & 22    & 22    & yes & 22 & 22 \\
PhantomEntryOnOrbit       & 1     & 1     & yes & 1 & 1 \\
UndisclosedTrackingLoss   & 1{,}104 & 1{,}104 & yes & \textbf{1{,}094} & 1{,}094 \\
DispositionDisagreement   & 932   & 932   & yes & \textbf{261} & \textbf{220} \\
CoverageGap               & 900   & 900   & yes & \textbf{622} & 622 \\
UnnumberedObject          & 605   & 605   & yes & 605 & 605 \\
IdentifierCollision       & 3     & 3     & yes & 3 & 3 \\
\bottomrule
\end{tabular}
\end{table}

The three errors have different characters, and only the first is the
common-mode failure this paper is about.

\paragraph{Disposition disagreement: 932, then 261, then 220.}
The count is the number of objects on which the two catalogues contradict each
other about whether the object is still in orbit. The 932 comprised 163 objects
GCAT records as gone while CelesTrak gives no decay date, and 769 objects
CelesTrak records as decayed while GCAT does not record them as gone. Producing
the count requires reading GCAT's status codes, and the pre-correction rule was
simple: a status in a set named \texttt{GONE} meant the object was gone, and
every other status meant it was still in orbit. That default did most of the
damage. Of the 769, 640 were objects whose GCAT record ends in a code that marks
a transition rather than a fate, such as docking, attachment, transfer or
renaming, and 500 of those were docking or attachment specifically. Another 57
ended in an explosion or a collision. The first correction removed 998
transition-coded objects from the comparison, 640 of which had been counted as
disagreements, and reclassified explosion and collision as destruction, giving
261. The ledger in Appendix~\ref{app:ledger} accounts for every object that
entered or left. Section~\ref{sec:second} shows that 261 was itself wrong.

\paragraph{Coverage gap: 900 against 622.}
The pipeline checked three of the four main catalogue files the source
publishes. The fourth was listed on the same index page and never fetched. It
holds 354 objects, 278 of which were being counted as absent from a catalogue
that contains them.

\paragraph{Undisclosed tracking loss: 1,104 against 1,094.}
Ten of the 1,104 objects do carry the CelesTrak field whose absence the count
is about. Correcting this strengthened the finding. The field is maintained and
set on 1,292 objects across the catalogue, ten of them among the 1,104, so the
claim moved from ``this register cannot record tracking loss'' to ``this
register records tracking loss and does not record it for 1,094 objects.''

None of the three errors is visible to a redundancy check, and only the first is
common-mode in the strict sense. What unites them is that each is an error about
what the source data means or contains, and the gate compares only what two
implementations derive from it.

\section{Why the gate could not see it}
\label{sec:mechanism}

\subsection{The diversity was in the wrong place}

The two paths differ in language, data structure and intermediate
representation. They do not differ in where their meaning comes from. The Python
path imports its classification constants and file loaders from the
reconciliation module. The SPARQL path queries a graph built by a script that
imports the same constants and loaders from the same module. We computed the
overlap from the source at the published commit: the names each path imports
from that module are identical, four vocabularies and two loaders
(Appendix~\ref{app:artifacts}, \texttt{shared\_ancestor\_audit.py}).

\begin{center}\small
\begin{tabular}{l}
\texttt{reconcile.py} defines \texttt{GONE}, \texttt{LEFT\_EARTH},
\texttt{ERROR}, \texttt{LOST}, \texttt{load\_gcat}, \texttt{load\_celestrak} \\
$\downarrow$ \hfill $\downarrow$ \\
\texttt{governance\_report.py} $\rightarrow$ Python counts \hfill
\texttt{build\_graph.py} $\rightarrow$ RDF graph $\rightarrow$ SPARQL counts \\
\end{tabular}
\end{center}

The gate diversified the last step of the computation and shared the first.
Whatever those sets assert about the source vocabulary is asserted identically on
both sides, so the agreement the gate reports is evidence that two pieces of
code, given the same premises, drew the same conclusion. In one sentence: the
gate verified that the Python and SPARQL paths implemented the same
misunderstanding identically.

\subsection{The defective definition}

The pre-correction disposition rule was one line applied to a set literal. An
object counted as gone in GCAT if any of its status codes was in

\begin{quote}\small\ttfamily
GONE = \{"R","D","L","LF","S","F","AF","AS","AR","AR IN","AL","AL IN","TX"\}
\end{quote}

\noindent and, through \texttt{bool(st \& GONE) != (n in ctdec)}, every other
object counted as still in orbit. The error lived in that default. Codes the set
did not name, among them \texttt{E} (exploded), \texttt{C} (collided),
\texttt{DK} (docked) and \texttt{ATT} (attached), were silently read as ``in
orbit''. A rule that made a claim only on explicit membership of \texttt{GONE}
or of the in-orbit set, and made no claim otherwise, returns 208 on the frozen
data, and all 208 are confirmed by the full-history check of
Section~\ref{sec:second}. The missing codes mattered far less than the decision
about what to do with codes nobody had classified.

A smaller defect in the same module shows the pathology from another angle. GCAT
marks uncertain dates with a trailing question mark, so a status can arrive as
\texttt{"R?"}. The original code compared raw strings and failed to match it,
which moved 15 objects in or out of the count. None of these corrections is
discoverable by comparing outputs. Each is a correction to a reading of the
source.

\subsection{Relation to multiversion programming}

Independently produced program versions do not fail independently
\citep{knight1986experimental}, and the accompanying theory located the cause in
the input rather than the versions: some inputs are hard for everyone, so
failures correlate however separately the versions were written
\citep{eckhardt1985theoretical, littlewood1989conceptual}. N-version designs were
proposed precisely to buy independence through separate development
\citep{avizienis1985nversion}. Our case differs in one structural respect that
makes it worse. In multiversion programming the shared artifact is a
specification, and versions can at least diverge in how they read it. Here the
shared artifact is an executable set of constants imported by both paths, so
there is no input on which they could disagree about a status code. The
correlation is not statistical but total.

\section{What did catch it}
\label{sec:caught}

The errors were found on a second pass over the same data, undertaken because
the results felt thin rather than because anything had flagged them. Three
checks did the work, and none compares two derivations of a quantity. Each goes
back to the source's own account of itself.

\paragraph{Read the field documentation, not the data.}
The vocabulary had been inferred from the values that appeared in the data,
which is how docking and attachment came to be read as statements about where an
object is. GCAT publishes a page defining every status value. It says that a
record represents a phase in an object's flight history and that the status
names the event that ends the phase. Reading that separated ``the object is
somewhere'' from ``this record stops here.''

\paragraph{Decompose every unexplained residue.}
The first write-up reported 769 of its disagreements as unexplained. Stating a
residue feels like rigour, which is why errors hide there comfortably. Asking what
the 769 consisted of ended most of the finding: 640 paired a CelesTrak impact or
landing with a GCAT transition code, 500 of them docking or attachment. Gemini~8
is the clearest case. GCAT ends the record at docking with the Agena target
vehicle and CelesTrak records the landing, and both are right about different
events in one mission.

\paragraph{Ask whether the source already models what you accuse it of omitting.}
The tracking-loss finding rested on an implicit claim that CelesTrak has no way
to record lost tracking. Checking that claim directly, rather than the count
derived from it, found the field in use.

Redundancy asks whether two routes from $A$ to $B$ agree. These three ask
whether $A$ is what we think it is.

\section{Checking the correction against object histories}
\label{sec:second}

The first correction moved \texttt{C} (collided) into the destroyed set, yet
GCAT's definitions allow an object to survive a collision. Checking that one
entry against the source found more than the entry itself.

\paragraph{The source's history of an object is in a file we never read.}
GCAT's definitions say a collision destroys the object ``in which case there is
no subsequent phase'', while a survivor starts a new phase; and that after
\texttt{E} (exploded) the ``next phase of this object is a debris fragment.''
So \texttt{E} marks a boundary, not a fate, and \texttt{C} is a fate only when
nothing follows it. Whether anything follows cannot be seen in the files the
pipeline read, because each holds exactly one record per object. GCAT's own
index says where the rest is: an event catalogue described as holding ``later
phases in history for objects in primary catalogs'', listed on the same page as
the four files we fetched, under a different heading.

\paragraph{Checking the correction against those histories.}
We appended each object's event-catalogue phases to its record, took the latest
phase as GCAT's current account of the object, resolved docked and attached
objects through the object they joined, and recomputed the comparison
(Appendix~\ref{app:history}). Of the 23 objects coded \texttt{C} that both
catalogues carry, 21 have a later phase; of the 78 coded \texttt{E}, 66 do. The
count falls from 261 to 220. It loses 42 objects, all 30 explosions and 12 of the
14 collisions in the 261. For 40 of them GCAT records a later phase that is still
in orbit, which is where CelesTrak has them; among these are Cosmos~2251 and
Iridium~33 after their 2009 collision and Fengyun-1C after the 2007
anti-satellite test. The other two are explosions with no later phase recorded,
on which GCAT makes no present claim. One object enters, a payload whose later
phase ends in a deorbit that CelesTrak does not record.

A second reading from the source agrees. GCAT publishes a derived catalogue of
the most recent phase of every object. Read against it, the count is 219, and 208
objects are common to both readings. Every one of the 23 differences has a named
cause: 12 are reentries recorded by one register and not yet by the other in the
six weeks between retrievals, 3 are GCAT record revisions in that window, and 8
are deep-space objects the pipeline sets aside by design and the derived
catalogue follows. Both readings drop all 42 objects. We treat 220 as the
reference count, with the caveat that its event-catalogue input was retrieved on
29~September (GCAT's data update of 28~September) rather than 18~August.

\paragraph{What this shows.}
The first correction was made by reading the documentation, as the paper
recommends, and it still misread two codes and missed the file that would have
settled them. It was a new reading of the vocabulary, and the pipeline's gate
and our three checks both inherited it. None of them
flagged the 42 objects (Section~\ref{sec:gates}); a later line-by-line reading
of the source's definitions did.

\section{Three gates, implemented and measured}
\label{sec:gates}

The checks in Section~\ref{sec:caught} were done by hand, and a discipline that
depends on someone deciding to look again is not a discipline. We implemented
all three and ran each on the defective code and on the first correction.
Constraint-based data validation checks declared expectations about data at
scale \citep{schelter2018deequ}; these gates instead check the pipeline's reading
of a vocabulary, which is where the defect lived. The second error gives a test
the first did not: the gates were built after the first error was known, but the
second was unknown when they were built, so how they behave on it is closer to a
prospective measurement. Table~\ref{tab:gates} gives the result.

\begin{table}[t]
\centering
\small
\caption{Each gate on the defective code, and on the first correction, which
still contained the second error of Section~\ref{sec:second} (42 objects).}
\label{tab:gates}
\begin{tabular}{>{\raggedright\arraybackslash}p{2.5cm}>{\raggedright\arraybackslash}p{3.9cm}>{\raggedright\arraybackslash}p{5.4cm}}
\toprule
Gate & Defective code & First correction, against the second error \\
\midrule
Total-function check on the status column &
\textbf{Fires}: 9 unclassified codes over 1{,}099 rows (\texttt{ATT},
\texttt{C}, \texttt{DK}, \texttt{E}, \texttt{EVA DP}, \texttt{GRP}, \texttt{N},
\texttt{REL}, \texttt{TFR}) &
\textbf{Silent}: 0 unclassified codes. \texttt{E} and \texttt{C} are classified,
wrongly \\
\addlinespace
Residue decomposition &
\textbf{Shows it}: 20 groups, top 5 hold 729 of 932; 500 pair docking or
attachment with impact or landing &
\textbf{Shows it, unread}: 44 of 261 in groups pairing \texttt{E} or \texttt{C}
with a CelesTrak object in orbit \\
\addlinespace
Source inventory reconciliation &
\textbf{Fires}: \texttt{satcat100k} listed, never fetched &
\textbf{Silent}: scoped to the four main catalogues; the event catalogue is on
the same index \\
\bottomrule
\end{tabular}
\end{table}

\paragraph{The total-function check tests completeness, not meaning.}
Requiring every observed status to be assigned to a named category attacks the
default that did the damage in Section~\ref{sec:mechanism}, and on the defective
code it lists every code implicated in the error. It cannot tell a right
assignment from a wrong one. The first correction assigned \texttt{E} and
\texttt{C} explicitly, to the wrong category, and the check passed. It also has a
scoping cost: applied to every controlled column rather than the one the
classifier consumes, it fires identically on both versions on 70,292 rows of
CelesTrak's orbit type, 70,130 of object type and 1,292 of data status, columns
the pipeline treats as opaque strings. The scope need not be chosen by hand. The
names both gate paths import from one module are exactly the definitions the
gate cannot test (Section~\ref{sec:mechanism}); at the published commit the
vocabularies among them are all classifications of the GCAT status column, which
is where the check separates cleanly.

\paragraph{Residue decomposition is an auditing aid, not a verdict.}
Grouping the unexplained disagreements by the two registers' fields made the
first error legible, since a status meaning ``joined another object'' should not
systematically meet a landing in the other register. It made the second error
legible too: on the corrected code, 30 explosions and 14 collisions sit against
CelesTrak objects in orbit. We read that table after the correction and saw
confirmation. The decomposition has no rejection criterion of its own, and the
reader supplied the wrong one.

\paragraph{Inventory reconciliation is exactly as good as its list.}
Comparing the files the source lists against the files the harvest loaded found
the missing fourth catalogue at once, with no false positive. On the second error
it was silent, because we had scoped it to the section of the index headed as
the main catalogues, while the index also lists the event catalogue, and eleven
object catalogues in all. Widening the list would have caught it and would also
flag files the pipeline has no reason to read. Running this gate also corrected
our own build report, which said the missing file recovered 280 objects; it
recovers 278, forced by the 900 and 622 that both reproduce exactly. A pass
undertaken to correct three published numbers introduced a fourth error.

\paragraph{A check that looked independent and was not.}
The SHACL validation returned 3{,}564 results, exactly the sum of six of the
seven counts. At the time this read as corroboration. It is an arithmetic
identity: the shapes select the nodes the counts count, from the graph the counts
were computed over, so when the counts were wrong the total was wrong by the same
amount and still matched. A suite of such steps produces a great deal of green
while measuring one thing repeatedly.

\section{Asking a model for an independent check}
\label{sec:automation}

The pipeline was written by people. We tested one part of the question for
automated authors: asked to add an independent second path for the disposition
count, does a code-generating model produce one whose meaning comes from
somewhere other than the path it was shown? Reuse is the expected answer, since
not duplicating a definition is correct practice by every other standard, and
generated test oracles are known to encode the behaviour of the code in front of
them rather than the behaviour intended \citep{konstantinou2024oracles}.

\paragraph{A preliminary run, and what was wrong with it.}
A preliminary run of 30 trials on 19~August used a coding-agent CLI with the
aliases \texttt{sonnet} and \texttt{haiku} and labelled outputs by pattern
matching over the generated source. It had four problems. The aliases were not resolved to model
identifiers in any record we kept. The CLI's default tools were not disabled, so
we cannot show that no trial read files. The prompt paraphrased the consumer
function: it counted only explicit membership of the gone and in-orbit sets,
where the real code treated every other status as in orbit. On the frozen data
that paraphrase returns 208, not 932, so the trials never saw the defect they
were scored against. And the loader bodies were elided, so no trial could see
GCAT's column names. What survives is narrower than we claimed. None of the 30
outputs introduced a status code absent from the prompt, and executed on the
real files, 22 of them return exactly the 208 of the path they were shown,
including all 19 that ran under the neutral and ``independent'' requests. When
code sharing was forbidden, three of ten reproduced 208 and seven diverged, none
by consulting the source: four guessed a column name and return 0, one parses
differently and returns 189, one crashes, and one reads GCAT's phase end date as
a decay date and returns 795 (Appendix~\ref{app:experiment}).

\paragraph{A controlled replication.}
We reran the experiment with the pre-correction code verbatim, loaders
included, so that the path shown returns the published 932. Each model was
pinned by identifier and called through the same CLI with tools disabled, a
one-line system prompt and an empty working directory: \texttt{claude-opus-5-5},
\texttt{claude-sonnet-5} and \texttt{claude-haiku-4-5-20251001}. Beside the three original requests,
two new conditions append GCAT's own status definitions verbatim, which is the
remedy this paper recommends. Outcomes are behavioural. Each generated path is
executed on a synthetic catalogue in which each pairing of a status code with a
CelesTrak state holds a distinct power of two of objects, so any count decodes to
exactly which pairings the path counted, and then on the frozen real files
(Appendix~\ref{app:experiment}).

\begin{table}[t]
\centering
\small
\caption{Controlled replication. Each cell is the number of generated paths that
return the published, defective 932 on the frozen data, out of the trials run.
``+ definitions'' appends GCAT's status definitions verbatim.
All 72 paths that return 932 count the same status pairings as the path they
were shown, on the probe or, for one, on the real files.}
\label{tab:reuse}
\begin{tabular}{lccc}
\toprule
Request & \texttt{opus-5-5} & \texttt{sonnet-5} & \texttt{haiku-4-5} \\
\midrule
Neutral                                  & 5/5 & 5/5 & 5/5 \\
``Independent'' path                     & 5/5 & 5/5 & 5/5 \\
Sharing forbidden                        & 4/5 & 4/5 & 5/5 \\
``Independent'' + definitions & 5/5 & 5/5 & 5/5 \\
Sharing forbidden + definitions & 4/5 & 5/5 & 5/5 \\
\bottomrule
\end{tabular}
\end{table}

Table~\ref{tab:reuse} gives the result. Of 75 trials, 72 return 932. Of the
other three, one \texttt{claude-sonnet-5} path returns 0 because its
hand-written file parser fails, and two \texttt{claude-opus-5-5} paths refuse
to produce a count. Only one of the 75 would have exposed the defect: with code
sharing forbidden, an Opus path wrote its own completeness check, and on the
real files it stops and lists the unclassified codes, among them \texttt{E},
\texttt{C}, \texttt{DK} and \texttt{ATT}. That is the first gate of
Section~\ref{sec:gates}, produced unprompted, once.

The definitions did not change what was computed. Of the 30 paths written with
GCAT's definitions in the prompt, 29 return 932, and every one of the 30 counts
\texttt{E}, \texttt{C}, \texttt{DK}, \texttt{ATT}, \texttt{TFR} and
\texttt{GRP} exactly as the shown path does wherever the probe can see it. The
paths read the documentation and kept the comparison. One Opus path annotated
\texttt{E} as ``exploded; the object continues as a debris fragment'' and
\texttt{C} as ``collided; may or may not survive, so not treated as gone'',
both correct, then placed both, with the docking codes, in a class of objects
``still in space or transition'' and compared that class with CelesTrak's decay
flag as the defective path does. On the real files its own completeness check
stops it on the undocumented suffixes \texttt{L?} and \texttt{R?}; on the
probe, which has none, it counts exactly what the defective path counts. A \texttt{claude-sonnet-5} path's docstring says
it ``re-derives the GCAT status taxonomy straight from the phase
documentation''; it returns 932.

\paragraph{What this does and does not establish.}
On this task, asking for an independent path, forbidding shared code, and
supplying the source's definitions each produced paths that, where they worked,
computed what the defective path computes, while looking to a code reviewer like
separate implementations with new names and comments citing the documentation.
It does not establish a rate for models in general (Section~\ref{sec:limits}). If a check is
meant to be independent, independence has to be specified as a property of its
inputs and verified by what the check computes, not requested; models also
struggle to correct their own reasoning without external feedback
\citep{huang2024cannotselfcorrect}.

\section{Limitations}
\label{sec:limits}

This is one incident in one pipeline, with one family of defects. We can show
that a redundancy gate failed this way here and name the structure that made it
fail, which is two paths importing one set of definitions, but we make no claim
about how often that structure occurs or how often it fails elsewhere. A rate
would need a corpus of pipelines with known ground truth, which is the scarce
thing. The same bound applies to the three gates. They are measured on one
codebase against two errors, and the second error, found after the gates existed and not by
them, is a single prospective test, not an evaluation.

The reference count is not an oracle. Neither register is treated as ground
truth, and 220 is the number of objects on which the two make incompatible
claims when GCAT is read through its own phase histories. Those histories come
from an event catalogue retrieved six weeks after the other files, filtered to
phases that began before the freeze; corrections GCAT made to older phases in
that window cannot be separated out. A second reading from GCAT's derived
catalogue gives 219 and agrees on 208 objects, with every difference assigned a
cause, and 8 of the differences are deep-space objects the comparison excludes
by design. Two explosions with no recorded later phase are set aside; counting
them as destroyed, as the first correction did, gives 222. The first correction
was presented as final, and a further pass could revise 220 as well.

The generation experiment is small: one task, three models from one family, five
trials per cell. A second family was planned and dropped, so nothing here speaks
to models outside it. No trial could retrieve anything: the definitions were
supplied in the prompt, and the event catalogue that settles \texttt{E} and
\texttt{C} was supplied to none, so the experiment does not test an agent that
can fetch the source for itself. We add one further limit: we are the party that benefits from the results being
true, which is a reason to weight the released prompts, raw outputs and
behavioural probe over our description of them.

\section{Conclusion}

A verification gate built on redundant computation reported agreement on seven
published counts, three of which were wrong, one overstated more than fourfold.
The gate was not defective. Its two paths differed in language, data structure
and representation, and they agreed because both imported one set of constants
encoding one misreading of a source catalogue's status vocabulary. Diversity
below a shared definition cannot detect an error in that definition, and the
more elaborate the redundancy, the more convincing the resulting green.

The case supports a design rule rather than a general law. The definitions two
gate paths share can be listed mechanically, and each of them needs a check that
goes back to the source rather than to another derivation. That rule is necessary
but not sufficient, as our own correction showed: it was made by reading the
documentation and still misread two codes, and the checks we built to prevent a
recurrence were silent on it. A further, line-by-line reading of the source's
definitions found it. Here redundant computation stood in only for a check of implementation, and
the errors that reached publication, twice, were errors of meaning.

\begin{ack}
We thank the anonymous reviewers and the area chair for their comments, and
Jonathan McDowell for maintaining GCAT and documenting it as thoroughly as he does.
\end{ack}

\bibliographystyle{plainnat}
\bibliography{refs}

\appendix

\section{Object-level ledger of the disposition count}
\label{app:ledger}

The unit throughout is an object present in both CelesTrak's catalogue and GCAT's
main catalogue, keyed by catalogue number (69,390 objects). Two
distinctions of unit matter when reading the figures below: 998 counted objects
removed from the comparison, not disagreements, and 500 is the docking and
attachment subset of the 769, not all of it. Table~\ref{tab:ledger} tracks every
object through the three versions; \texttt{count\_ledger.py} produces each
figure and asserts that the arithmetic closes.

\begin{table}[h]
\centering
\small
\caption{The disposition count, object by object. Status groups refer to the
object's GCAT status code.}
\label{tab:ledger}
\begin{tabular}{lr}
\toprule
\textbf{Published count} & \textbf{932} \\
\quad GCAT gone, CelesTrak gives no decay date & 163 \\
\quad CelesTrak decayed, GCAT not gone & 769 \\
\qquad docking or attachment (\texttt{DK}, \texttt{ATT}) & 500 \\
\qquad other transition codes (\texttt{TFR}, \texttt{GRP}, \texttt{N}, \ldots) & 140 \\
\qquad explosion or collision (\texttt{E}, \texttt{C}) & 57 \\
\qquad in orbit but tracking lost (\texttt{OX}) & 31 \\
\qquad error entries (\texttt{ERR}) & 21 \\
\qquad in orbit (\texttt{O}, \texttt{AO}) & 14 \\
\qquad gone, but status carries a trailing \texttt{?} & 6 \\
\midrule
Left the comparison at the first correction & $-724$ \\
\quad transition codes now excluded (of 998 objects excluded in all) & $-640$ \\
\quad explosions and collisions reclassified as destroyed, CelesTrak agrees & $-57$ \\
\quad error entries now excluded & $-21$ \\
\quad trailing \texttt{?} now stripped & $-6$ \\
Entered at the first correction & $+53$ \\
\quad explosions and collisions reclassified as destroyed, CelesTrak silent & $+44$ \\
\quad trailing \texttt{?} now stripped & $+9$ \\
\textbf{First correction} & \textbf{261} \\
\midrule
Left at the full-history check: later phase in orbit or unrecorded & $-42$ \\
\quad explosions & $-30$ \\
\quad collisions & $-12$ \\
Entered: attached object whose later phase ends in a deorbit & $+1$ \\
\textbf{Full-history count} & \textbf{220} \\
\bottomrule
\end{tabular}
\end{table}

\section{The full-history check}
\label{app:history}

\texttt{history\_check.py} appends to each object's record the phases GCAT's
event catalogue lists for the same identifier, drops the 19 phases that began
after 18~August~2026, and takes the status of the latest phase. Following GCAT's
definitions, \texttt{C} counts as destruction only as a last phase; \texttt{E},
\texttt{N}, \texttt{LEASE} and \texttt{REFLT} as a last phase leave the present
state unrecorded (28 objects, set aside); docking, grappling, attachment and
transfer codes resolve through the object named as destination. The three
catalogue numbers GCAT assigns to two objects each are resolved on both, and
count only if both agree. The 14 objects whose last phase is a lease are all in
orbit in CelesTrak, so treating a lease as a free-flying state instead leaves the
count unchanged.

\texttt{currentcat\_crosscheck.py} repeats the comparison from GCAT's derived
catalogue of the most recent phase of each object, mapping every status text
that occurs to a disposition and failing on any it cannot map. Over 69,249
comparable objects it gives 219, with 208 in common with the 220. The 12 objects
only in our reading are 9 reentries CelesTrak recorded between 13 and 18~August
that the frozen GCAT file did not yet carry, 2 collisions with no later phase
that the derived catalogue now labels as explosions, and 1 record GCAT revised
after the freeze. The 11 only in the derived reading are 8 deep-space objects,
which the pipeline sets aside, and 3 reentries GCAT recorded after the freeze.
Of the 42 objects removed from the first correction, the derived catalogue lists
40 as in Earth orbit and 2 as exploded; CelesTrak gives none of the 42 a decay
date.

\section{Generation experiment: protocol and per-trial results}
\label{app:experiment}

\paragraph{Behavioural probe.}
\texttt{behaviour\_probe.py} writes a synthetic CelesTrak file and GCAT file
under the paths the prompt names. There are 14 cells: each of \texttt{E},
\texttt{C}, \texttt{DK}, \texttt{ATT}, \texttt{TFR} and \texttt{GRP}, paired
with a CelesTrak decay date and without one, plus two controls every reading
counts (\texttt{R} with no decay date, \texttt{O} with one). Cell $i$ holds $2^i$
objects, all with five-digit catalogue numbers so zero-padded and bare forms
coincide. The generated code runs in a separate process with the prompt's module
importable and its names preloaded, since many paths use them without importing
them; the original path is stubbed to return a sentinel that decodes to nothing.
Every zero-argument function the code defines, other than loaders and helpers,
is called, and any returned or printed count that decodes to a set containing
both controls is taken as the path's signature. The same code is then run on the
frozen real files and the number each chosen function returns is recorded.
Reference signatures are obtained by running the reference code through the
same harness, not written down; the verbatim pre-correction path returns 932 on
the real files, which checks the harness end to end.

\paragraph{Preliminary run (19 August).}
Prompt in \texttt{reuse\_experiment.py}; raw outputs in \texttt{reuse\_raw/};
labels in \texttt{behaviour\_labels.json}. The path shown counts only the two
controls on the probe and returns 208 on the real files. Neutral request: 9 of 10
reproduce 208; 1 does not run (it imports a name the module does not define).
``Independent'': 10 of 10 reproduce 208. Sharing forbidden: 3 reproduce 208; 4
read GCAT's catalogue-number column under a guessed name and return 0; 1 parses
the files differently and returns 189; 1 crashes on real rows; 1 ignores the
status codes, treats any GCAT phase end date as a decay, and returns 795. The
earlier regex labels called all 30 ``inherits the defect''; that label
described the vocabulary each path contains, not what it computes.

\paragraph{Replication (29 September).}
Runner \texttt{reuse\_replication.py}; prompts, raw outputs and per-trial
metadata in \texttt{replication\_raw/}. Every model was called through the
Claude Code CLI with \texttt{--tools ""},
\texttt{--system-prompt "You are a helpful assistant."},
\texttt{--setting-sources ""}, \texttt{--strict-mcp-config} and no session
persistence, from an empty directory. Sonnet and Haiku ran on CLI version
2.1.277; \texttt{claude-opus-5-5} needs 2.1.280 or later and ran on 2.1.284. The
CLI does not expose sampling parameters, and its defaults include extended
thinking, which every trial used. It also makes a small fixed auxiliary call to a
Haiku model; the recorded per-model token usage shows that in every trial the
answer's output tokens came from the requested model, in a single turn with no
tool use. A local model from a second family was started and abandoned because
it decoded too slowly to complete the design; no partial result is reported.

\paragraph{Per-trial outcomes.}
On the probe, 72 of the 75 paths have exactly the signature of the path shown,
which counts all six codes as in orbit whenever CelesTrak records a decay. The
three exceptions: one \texttt{claude-sonnet-5} path (sharing forbidden) returns 0
on probe and real files because its hand-written CSV parser fails; one
\texttt{claude-sonnet-5} path (sharing forbidden, with definitions) takes its
decay signal from CelesTrak's operational status field, which the probe leaves
blank, and returns 932 on the real files; one \texttt{claude-opus-5-5} path
(sharing forbidden) raises on every status it has not classified, which on the
real files includes \texttt{ATT}, \texttt{C}, \texttt{DK}, \texttt{E},
\texttt{EVA DP} and \texttt{GRP}. A second Opus path (sharing forbidden, with
definitions) matches the shown path on the probe and on the real files stops on
the undocumented \texttt{L?} and \texttt{R?}. Three \texttt{claude-haiku-4-5}
paths and several Opus paths reference a constant of the prompt's module without
importing it or run as command-line programs; the harness supplies the constant
and a data-directory argument, as the pipeline would, and reads counts that are
printed rather than returned.

\section{Artifacts}
\label{app:artifacts}

Code, gate implementations, experiment prompts, raw model outputs, per-trial
metadata and the behavioural probe are in the project repository at
\url{https://github.com/fabio-rovai/space-object-register-ontology} under
\texttt{paper/gates/}. The source files are public and were retrieved on the dates
below; the SHA-256 prefixes identify the exact copies used.

\begin{center}\small
\begin{tabular}{lll}
\toprule
File & Retrieved & SHA-256 (prefix) \\
\midrule
CelesTrak \texttt{satcat.csv} & 18 Aug 2026 & \texttt{888f67c53ee3d006} \\
GCAT \texttt{satcat.tsv} & 18 Aug 2026 & \texttt{33751fb8ac5b6fb1} \\
GCAT \texttt{auxcat.tsv} & 18 Aug 2026 & \texttt{82363eb712d8a392} \\
GCAT \texttt{ftocat.tsv} & 18 Aug 2026 & \texttt{d6b6351ceb0a5a98} \\
GCAT \texttt{satcat100k.tsv} & 18 Aug 2026 & \texttt{6ec65d36db9f6263} \\
GCAT \texttt{ecat.tsv} (event catalogue) & 29 Sep 2026 & \texttt{39023a287c27326e} \\
GCAT \texttt{currentcat.tsv} (derived) & 29 Sep 2026 & \texttt{8a0d9285b6fce7b1} \\
\bottomrule
\end{tabular}
\end{center}

\end{document}